\documentclass{article}
\usepackage{spconf,amsmath,graphicx,booktabs,multirow,amssymb,flexisym,pdfpages}

\title{Coincidence, Categorization, and Consolidation: \\ Learning to Recognize
  Sounds with Minimal Supervision}
\makeatletter
\def\@name{ \emph{Aren Jansen, Daniel P. W. Ellis, Shawn Hershey,}
  \\\thanks{This extended version of a ICASSP 2020 submission under same title
    has an added figure and additional discussion for easier consumption.}
    \emph{R. Channing Moore, Manoj Plakal, Ashok C. Popat, Rif
      A. Saurous}\vspace{6pt}}
\makeatother
\address{Google Research, Mountain View, CA, and New York, NY, USA\\
{\footnotesize \tt\{arenjansen,dpwe,shershey,channingmoore,plakal,popat,rif\}@google.com}}

\begin{document}
\ninept
\maketitle
\begin{abstract}
Humans do not acquire perceptual abilities in the way we train machines.  While
machine learning algorithms typically operate on large collections of
randomly-chosen, explicitly-labeled examples, human acquisition relies more
heavily on multimodal unsupervised learning (as infants) and active learning (as
children).  With this motivation, we present a learning framework for sound
representation and recognition that combines (i) a self-supervised objective
based on a general notion of unimodal and cross-modal coincidence, (ii) a
clustering objective that reflects our need to impose categorical structure on
our experiences, and (iii) a cluster-based active learning procedure that
solicits targeted weak supervision to consolidate categories into relevant
semantic classes.  By training a combined sound
embedding/clustering/classification network according to these criteria, we
achieve a new state-of-the-art unsupervised audio representation and demonstrate
up to a 20-fold reduction in the number of labels required to reach a desired
classification performance.
\end{abstract}


\section{Introduction}
\label{sec:intro}

In the first year of life, typical infants are awake for $\sim$4000 hours,
during which they are presented with a wide variety of environmental sounds,
infant-directed speech, and a companion visual stream of over 1M images
(assuming 1 fps). It is only after this pre-verbal exposure that our abilities
of object tracking, color discrimination, object recognition, word and phoneme
recognition, and environmental sound recognition
emerge~\cite{law2011infant,sugita2004experience,braddick2009infants,mandel1995infants,dupoux2018cognitive,cummings2009infants}.
Beginning in the second year, children become proficient at knowing what they do
not know and solicit explicit labels for novel classes of stimuli they encounter
using finger pointing and direct
questions~\cite{goupil2016infants,clark2009first,ganger2004reexamining,begus2012infant}.
However, this process is not carried out on an per-instance basis; instead,
children exploit their past learning of invariances (e.g. rotation/lighting for
vision, speaker for speech, loudness for sounds) to generalize a single label
from a caregiver to a range of stimuli.

These aspects of early human learning are not captured by the traditional
supervised learning practice of collecting a large set of explicitly labeled
examples and using it to train a model from scratch.  Instead, it is clear
humans also rely on some combination of unimodal/multimodal unsupervised
learning and active learning to acquire these abilities.  With this inspiration,
this paper presents a joint learning framework that unites several strands of
unsupervised deep learning research to train high quality semantic sound models.
As input, we are provided a large collection of unlabeled video data and the
goal is to require only a minimal amount of manual, targeted annotation.  Our
framework rests on three learning mechanisms: (i) observing coincidences both
within and across modalities, (ii) discovering and imposing categorical
structure, and (iii) consolidating those categories into practical semantic
classes.

Unsupervised learning has undergone major advances with the development of
so-called self-supervised learning methods, which define application-specific
proxy tasks to encourage neural networks to produce semantically structured
representations.  We propose a general unimodal and cross-modal representation
learning technique based on the proxy task of coincidence prediction, which
unifies recent work in audio-only~\cite{jansen2018unsupervised} and
audio-visual~\cite{arandjelovic2017look,cramer2019look} self-supervised
learning.  The goal is to learn separate audio and image embeddings that can
predict whether each sound-sound pair or each sound-image pair occurs within
some prescribed temporal proximity in which semantic constituents are generally
stable.  Each of these two prediction tasks elicits strong semantic encoding and
we demonstrate further improvement through their combination.

Once we have a semantically structured representation, we can initiate the
discovery of categorical structures that ultimately facilitate connecting our
perception to higher-level cognitive reasoning.  While any traditional
clustering algorithm might apply, we propose a novel, neural network-based
clustering procedure that not only provides a partition of the embedding space
but also updates the coincidence-trained embedding network to reinforce
discovered categories.  We demonstrate that this procedure provides further
improvements to our embedding model, resulting in a new state-of-the-art
unsupervised audio representation.

Finally, automatically discovered categories are not particularly useful until
grounded to a prescribed ontology.  Traditional active learning methods require
access to a pre-existing classifier for prioritization.  Absent such a
classifier in truly unsupervised settings, we instead adopt a cluster-based
active learning procedure~\cite{urner2013plal,jansen2017large} whereby we weakly
label each discovered category by soliciting an annotation for a randomly
selected sound contained within.  This neutralizes the sampling hazards of
skewed data distributions while functioning to consolidate overly-specialized
discovered categories into more meaningful and generalizable classes.  Using
this targeted annotation procedure, we can obtain weak labels for nearly the
entire dataset, which we use to train sound classifiers that are initialized
with the unsupervised semantic representation network.  The end result is
dramatic improvements in classification performance in the usual case where
access to explicit supervision is limited.

\section{Related Work}
\label{sec:related}

A wide variety of self-supervised methods have been developed in the audio and
computer vision communities.  For sound events, temporal proximity, temporal lag
prediction, context reconstruction, and sound mixing have been demonstrated to
be effective~\cite{jansen2018unsupervised,tagliasacchi2019self}.  In computer
vision, proxy objectives have been based on
egomotion~\cite{agrawal2015learning}, spatial/compositional
coherence~\cite{doersch2015unsupervised,pathak2016context,oord2018representation},
temporal coherence/proximity in
video~\cite{mobahi2009deep,redondo2019unsupervised}, object tracking in
video~\cite{wang2015unsupervised}, colorization~\cite{zhang2016colorful}, and
rotation~\cite{gidaris2018unsupervised}.  The coincidence-based approach we
propose directly captures these temporal coherence and proximity criteria.
Furthermore, for a sufficiently rich video training dataset in which lighting,
camera angle, and object position are all dynamic, coincidence can subsume most
of the other listed image-based self-supervised objectives as well.

Recognizing the limitations of unimodal self-supervised methods, researchers
have increasingly focused on multimodal training objectives that introduce
powerful cross-modal constraints. Three prominent deep learning approaches are
(i) Deep Canonical Correlation Analysis (DeepCCA)~\cite{andrew2013deep}, a deep
learning generalization of linear CCA; (ii) “Look, Listen, and Learn”
($L^3$)~\cite{arandjelovic2017look}, which learns representations that can
predict audio-visual frame correspondence in videos; and (iii) metric learning
losses applied to shared multimodal embedding
spaces~\cite{harwath2018jointly,zhao2018sound,senocak2018learning,chung2016out}. In
all three cases, the learning is driven by the powerful cue of coincidence of
semantic content present in parallel observations across the modalities.

Cluster-based active learning in fixed representation spaces has been proposed
in past research~\cite{urner2013plal,jansen2017large}.  However, to the best of
our knowledge, ours is the first attempt to train a neural network with a
cluster-based objective to facilitate an active learning procedure.  There have
also been recent attempts at neural-based
clustering~\cite{caron2018deep,arora2019deep,xie2016unsupervised}.  However,
each one of those methods reinforces classical k-means solutions, while our
proposed approach performs the clustering from scratch, entirely within the
context of neural network training.

\begin{figure}
  \centering \includegraphics[width=1.0\columnwidth]{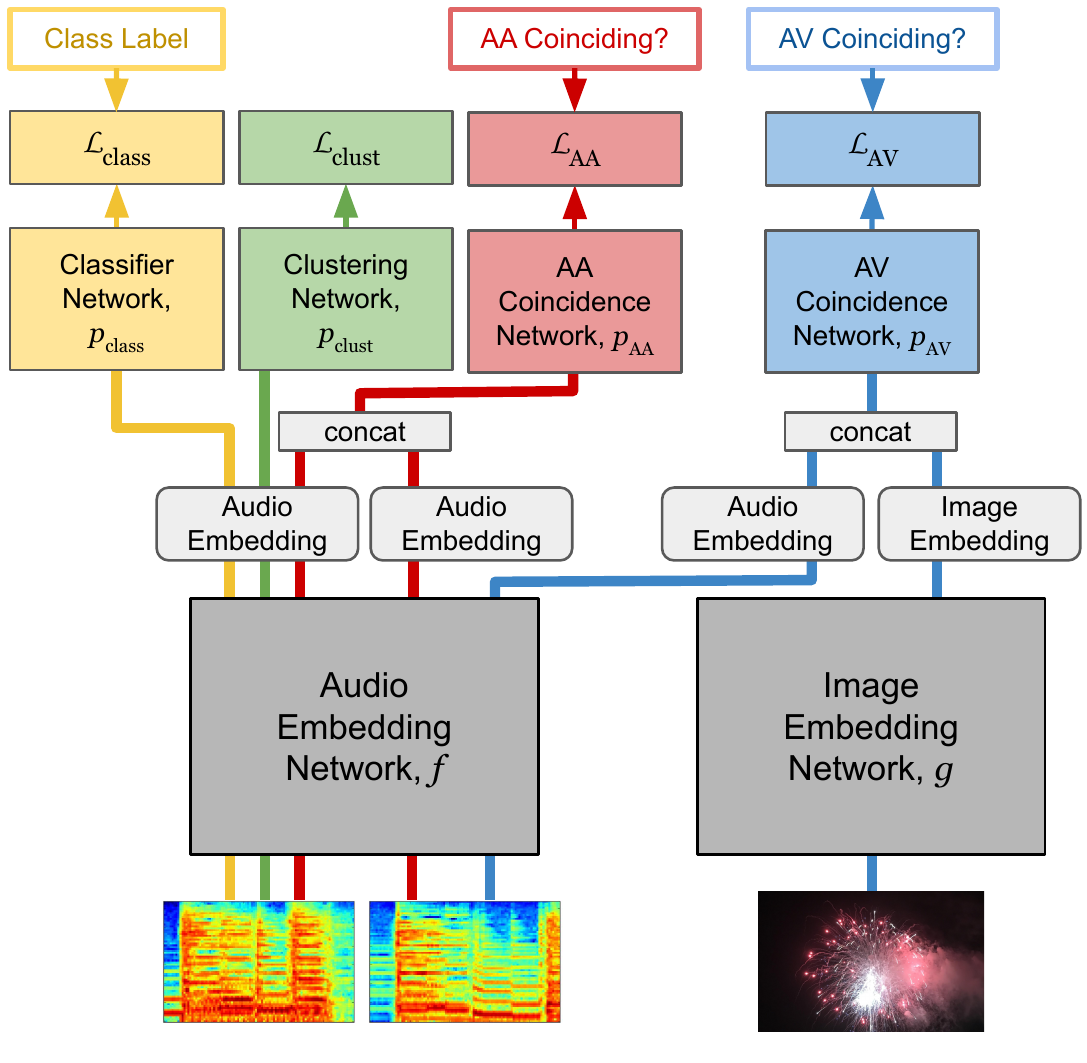}
  \caption{Learning framework diagram. Each of the four loss function processing
    paths is specified with a color: red for audio-audio (AA) coincidence,
    blue for audio-visual (AV) coincidence, green for clustering, and yellow for
    classification.}
  \label{fig:schem}
\end{figure}

\section{The Learning Framework}
\label{sec:framework}

Our goal is to train a single deep audio embedding network that defines a map
$f\!:\!\mathbb{R}^{F \times T} \rightarrow \mathbb{R}^d$ from sounds (represented
by log mel spectrogram context windows with $F$ frequency channels and $T$
frames) to a $d$-dimensional representation that supports sound recognition and
retrieval tasks.  We use a combination of three learning mechanisms, each
involving their own auxiliary networks and losses, described in turn below.

\subsection{Generalized Coincidence Prediction}
\label{sec:coincidence}

Our coincidence prediction approach is based on the assumption that the set of
semantic categories we interact with changes much more slowly than the raw
pixels and sounds we perceive on an instant-by-instant basis.  Therefore, there
must exist a relatively stable latent representation of the raw inputs that
reflects semantic content and ignores higher frequency extrinsic variation.
Such a representation would facilitate prediction of whether a pair of inputs
are coinciding, since temporal proximity would be correlated with semantic
similarity.  Coincidence prediction then becomes a proxy task for semantic
representation learning.  Critical to this proxy task is the choice of a
suitable time scale of coincidence, which we denote $\Delta T$.  The appropriate
value is task-dependent and needs to correspond roughly to the rate of semantic
change in the input.  Our prior work in temporal proximity-based metric
learning~\cite{jansen2018unsupervised} was a direct attempt to leverage
coincidence for audio representation learning. There, we learned a
representation in which spatial proximity was correlated with temporal proximity
($\Delta T = 10$ s) of the inputs, which elicited useful semantic structure in
the embedding space.  Our present goal is to extend that approach into a single
methodology that can also exploit cross-modal audio-visual self-supervision. 

Our coincidence prediction approach is a generalization of the audio-visual (AV)
correspondence proxy task~\cite{arandjelovic2017look}. That work's key
innovation was to simultaneously train an audio and image embedding network on
unlabeled video that supported prediction of whether an audio and image frame
occurred at the same (corresponding) point in time.  We introduce three
modifications to the that recipe.  First, since it is clear from human
experience that we need not see an object make a sound to associate them
semantically, we relax the time scale in what we call a less restrictive
coincidence prediction.  Second, we use the same coincidence prediction strategy
to exploit both unimodal (audio-audio, for which correspondence is not useful)
and cross-modal input pairs to train a single shared audio embedding model.
Finally, we improve optimization (and ultimately performance) by exploiting all
non-coincident pairs in each minibatch (rather than a single random selection).

As depicted in the red path of Figure~\ref{fig:schem}, our audio-audio (AA)
coincidence prediction task is trained on a large collection of
coincidence-labeled audio example pairs of the form $(x_1, x_2, y)$, where each
$x_i \in \mathbb{R}^{F \times T}$, and $y \in \{0, 1\}$ is a coincidence
indicator for a time scale $\Delta T$.  Each audio example is passed through the
audio embedding network $f$, and the outputs are concatenated into a vector $z =
\left[f(x_1), f(x_2)\right] \in \mathbb{R}^{2d}$.  The AA coincidence prediction
task is performed by a fully connected binary classification network
$p_{\mathrm{AA}} \!:\! \mathbb{R}^{2d} \rightarrow [0, 1]$ that maps each $z$
into the probability that the input pair was coinciding.  Given a batch of
coinciding example pairs $X = \{(x_1^{(i)}, x_2^{(i)})\}_{i=1}^B$, we construct
$B \cdot (B-1)$ negative examples by assuming each pair $(x_1^{(i)}, x_2^{(j)})$
for $i \!\neq\! j$ are non-coinciding.  This all-pairs construction introduces
negligible label noise and uses each mini-batch completely without having to
resort to within mini-batch mining techniques required by triplet loss
methods~\cite{jansen2018unsupervised}.  The resulting AA coincidence loss
function is the balanced cross-entropy, given by

\begin{align}
  \begin{split}
  \mathcal{L}_{\mathrm{AA}} & (X) = 
  -\frac{1}{B}\sum_{i=1}^B \log p_{\mathrm{AA}}([f(x_1^{(i)}), f(x_2^{(i)})]) \\
  & - \frac{1}{B(B\!-\!1)}\sum\limits_{\substack{1 \le i,j \le B \\ j \ne i}}
  \log \left[ 1\!-\!p_{\mathrm{AA}}([f(x_1^{(i)}), f(x_2^{(j)}])) \right].
  \end{split}
  \label{eq:LAA}
\end{align}

\noindent
The AV coincidence prediction task (blue path in Figure~\ref{fig:schem})
operates similarly with three differences.  First, each coincidence-labeled
training pair $(x_1, x_2, y)$ now has $x_2 \in \mathbb{R}^{W\!\times\!
  H\!\times\! D}$, each $W \!\times\! H$ pixel images color depth $D$.  Second,
the image inputs are processed by their own embedding network $g \!:\!
\mathbb{R}^{W\!\times\! H\! \times\! D} \rightarrow \mathbb{R}^d$, which is
jointly trained alongside $f$. Third, we introduce a dedicated second AV
coincidence prediction network $p_{\mathrm{AV}}$. The AV coincidence loss
function takes the form 

\begin{align}
  \begin{split}
    \mathcal{L}_{\mathrm{AV}} & (X) =
    -\frac{1}{B}\sum_{i=1}^B \log p_{\mathrm{AV}}([f(x_1^{(i)}), g(x_2^{(i)})]) \\
    & - \frac{1}{B(B\!-\!1)}\sum\limits_{\substack{1 \le i,j \le B \\ j \ne i}}
    \log \left[ 1\!-\!p_{\mathrm{AV}}([f(x_1^{(i)}), g(x_2^{(j)}])) \right].
  \end{split}
  \label{eq:LAV}
\end{align}

\noindent
This is the same loss form of Equation~\ref{eq:LAA}, but with $p_{\mathrm{AA}}$
and $f(x_2)$ replaced with $p_{\mathrm{AV}}$ and $g(x_2)$, respectively.

\subsection{Categorization with Entropy-based Clustering}
\label{sec:categorization}

We posit that a good clustering is a partition of data points such that (i) each
data point is confidently assigned to one cluster, (ii) all available clusters
are used, and (iii) the set of points assigned to the same clusters are close
under some relevant metric (Euclidean or otherwise).  To learn such a $K$-way
partition, we introduce a map $p_{\mathrm{clust}} \!:\!  \mathbb{R}^d \rightarrow
[0,1]^K$ from our embedding space to a categorical distribution specifying the
probability of assignment to each cluster (see the green path in
Figure~\ref{fig:schem}).  We can encourage confident assignment by reducing
per-data-point entropy of $p_{\mathrm{clust}}$.  However, to prevent the trivial
solution that assigns all points to one cluster, we must add a countervailing
objective that increases the entropy of the $p_{\mathrm{clust}}$ distribution
averaged over the whole dataset.  Finally, by expressing the map
$p_{\mathrm{clust}}$ with a neural network of limited complexity, we ensure
preservation of locality.

These objectives are amenable to stochastic gradient descent optimization using
audio example mini-batches $X=\{x_i\}_{i=1}^B$, where each $x_i \in
\mathbb{R}^{F \times T}$, and the loss function

\begin{equation*}
  \mathcal{L}_{\mathrm{clust}}(X) =
  \frac{1}{B} \sum_{i=1}^B H[p_{\mathrm{clust}}(f(x_i))] -
  \gamma H\left[\frac{1}{B} \sum_{i=1}^B p_{\mathrm{clust}}(f(x_i))\right],
\end{equation*}

\noindent
where $f$ is the audio embedding map defined above, $H[\cdot]$ denotes entropy,
and $\gamma$ is a diversity hyperparameter.  Increasing $\gamma$ encourages a
more uniform cluster occupancy distribution and, given sufficiently large $K$,
is the primary setting for model selection.  For the special case of $\gamma=1$,
minimizing $\mathcal{L}_{\mathrm{clust}}$ reduces to maximizing the mutual
information (MI) between the model inputs and the output clusters, which was
previously introduced as a discriminative clustering
objective~\cite{krause2010discriminative}. As that work indicated, MI
maximization alone finds trivial solutions, which they address with explicit
regularization terms, but they still required k-means initialization for
successful application.  Setting our hyperparameter $\gamma > 1$ also acts to
prevent trivial solutions.  Critically, however, our objective is amenable to
cold start training and can be used to fine tune embedding networks, where the
representation evolves during optimization.

\subsection{Consolidation with Cluster-based Active Learning}
\label{sec:consolidation}

Given an imperfect semantic representation, each class will be subject to some
degree of fragmentation into multiple modes, while some classes will fail to
split into separable structures.  Thus, one tenable categorization strategy is
to over-partition the space such that hypothesized units remain pure and
over-specialized.  In this case the final requirement of learning semantic
classes is a consolidation, or grouping, of the discovered categories into
broader, more general units via explicit supervision.  However, this explicit
supervision need not be provided for every example in every cluster.  Instead,
if the clusters are sufficiently pure, we can simply request a label for a
single, randomly-selected cluster constituent and propagate that label to the
cluster cohabitants.  This strategy defines an active learning procedure that
requires no pre-existing classifier.

There is natural trade-off between cluster purity and semantic fragmentation in
the discovery of categorical structure in a representational space.  On one
extreme, each data point can be assigned its own category, achieving perfect
purity, but with maximal fragmentation.  On the other extreme, where all points
are placed into one bin, all examples for each class are colocated, but there is
no semantic discrimination whatsoever.  In the context of a cluster-based active
learning procedure, the concepts of purity and fragmentation translate into
resulting label noise (precision) and label quantity (recall) for each given
cluster labeling budget, a trade-off we explore.  Note we found that alternative
schemes of labeling more than one example per cluster (and thus, fewer clusters)
were not as effective as simply labeling more clusters with the same budget.

Once we have performed this labeling procedure, we will have a collection of
labeled examples, which we split into batches of the form $Z =
\{(x_i,y_i)\}_{i=1}^B$, where each $x_i \in \mathbb{R}^{F \times T}$ and $y_i
\in \{0,1\}^C$ for a $C$-way classification task (see the yellow path in
Figure~\ref{fig:schem}) .  We can then define our training objective as 

\begin{equation}
  \mathcal{L}_{\mathrm{class}}(Z) =
  \frac{1}{B} \sum_{i=1}^B H[y_i,p_{\mathrm{class}}(f(x_i))],
  \label{eq:class}
\end{equation}

\noindent
where $p_{\mathrm{class}}: \mathbb{R}^d \rightarrow [0,1]^C$ is the $C$-way
output distribution of a classification network operating on the learned audio
embeddings and $H[\cdot,\cdot]$ denotes the cross entropy between labels and
predictions. 

Finally, it is worth noting that most unsupervised representation learning
studies evaluate the utility of the learned embeddings in a lightly supervised
classification evaluation, which assumes a small sample of labeled examples with
relatively uniform class representation.  However, these studies never account
for where that small labeled dataset came from.  If provided an unbiased sample
for annotation, natural class skew will mean oversampling of common classes and
the potential to miss some classes entirely.  The cluster-based active learning
procedure described above is a natural solution to this problem as well.

\subsection{Learning Curriculum}
\label{sec:curriculum}

Jointly optimizing the four objectives listed above, each of which involves
specialized auxiliary networks, proves challenging with stochastic gradient
descent optimization.  Therefore, we devised a staged learning curriculum that
applies the objectives in sequence, first with unsupervised losses in descending
order of learning signal, followed by the supervised loss to produce a
classifier after labeling.  Specifically, we begin by minimizing
$\mathcal{L}_{\mathrm{AV}}$ of Equation~\ref{eq:LAV} until convergence. Next, we
continue by minimizing the joint AV and AA coincidence loss

\begin{equation*}
  \mathcal{L}_{\mathrm{coin}} = (1-\alpha) \mathcal{L}_{\mathrm{AV}} + \alpha
  \mathcal{L}_{\mathrm{AA}},
\end{equation*}

\noindent
where $\alpha \in [0,1]$ is an interpolation constant hyperparameters and
$\mathcal{L}_{\mathrm{AA}}$ is given by Equation~\ref{eq:LAA}. We then introduce
the clustering objective and minimize the joint unsupervised loss

\begin{equation*}
\mathcal{L}_{\mathrm{joint}} = (1-\beta) \mathcal{L}_{\mathrm{coin}} +
\beta \mathcal{L}_{\mathrm{clust}},
\end{equation*}

\noindent
where $\beta \in [0,1]$ is another interpolation hyperparameter.  Finally, after
cluster-based labeling, we fine-tune the embedding model, $f$, using only the
classifier objective, $\mathcal{L}_{\mathrm{class}}$ of Equation~\ref{eq:class}. 

\begin{table*}[t]
\vspace{-0.2cm}
  \centering
  \caption{Performance of segment retrieval and shallow model
    classification with fixed representations. All embedding models use
    ResNet-50 with $d=128$, $\Delta T$=10 s, and all-pairs batches (unless
    noted).}
  \label{tab:comb1}
  \vspace{0.2cm}
  \begin{tabular}{l|c|c|c|c}
    \hline
    & \multicolumn{2}{c|}{{\bf QbE Retrieval}} & \multicolumn{2}{c}{{\bf Classification}}\\
    \cline{2-5}
    {\bf Representation} & {\bf mAP} & {\bf recovery} & {\bf mAP} & {\bf recovery} \\
    \hline
    a. Explicit Label Triplet (topline)
    & 0.784 & \emph{100\%} & 0.288 & \emph{100\%} \\
    b. Log Mel Spectrogram (baseline)
    & 0.421 & \emph{0\%}   & 0.065 & \emph{0\%} \\
    \hline
    c. Temporal Proximity Triplet~\cite{jansen2018unsupervised}
    & 0.549 & 35\%         & 0.226 & 72\% \\
    d. AV Correspondence (VGG)~\cite{arandjelovic2017look}
    & 0.625 & 56\%         & 0.249 & 83\% \\
    \hline
    e. AA Coincidence $(\mathcal{L}_{\mathrm{AA}})$
    & 0.552 & 36\%         & 0.206 & 63\% \\
    f. AV Coincidence $(\mathcal{L}_{\mathrm{AV}})$
    & 0.669 & 68\%         & 0.269 & 92\% \\
    \mbox{        } $\boldsymbol{\cdot}$ g. ResNet $\rightarrow$ VGG
    & 0.629 & 57\%         & 0.265 & 90\% \\
    \mbox{        } $\boldsymbol{\cdot}$ h. All-pairs $\rightarrow$ random negatives 
    & 0.641 & 61\%         & 0.253 & 84\% \\
    \mbox{        } $\boldsymbol{\cdot}$ i. $\Delta T\! =\!10$ s $\rightarrow 1$ s
    & 0.659 & 66\%         & 0.270 & 92\% \\

    \hline
    j. AA+AV Coincidence $(\mathcal{L}_{\mathrm{coin}})$
    & 0.677 & 71\%         & 0.282 & 97\% \\
    k. AA+AV Coincidence + Cluster:$K\!=\!$ 1M $(\mathcal{L}_{\mathrm{joint}})$
    & 0.705 & 78\%         & 0.285 & 99\% \\
    \hline
  \end{tabular}
\end{table*}

\section{Experiments}
\label{sec:exp}

We evaluate our proposed learning framework using the AudioSet
dataset~\cite{audioset}, consisting of over 2 million video clips, each
approximately 10 seconds in duration and labeled using an ontology of 527 audio
classes.  In contrast to most past studies that use the dataset, we use both the
audio and video, sampling at 1 Hz both the image frames (scaled to
$W\!=\!H\!=\!128$ pixels and color depth $D\!=\!3$) and log mel spectrogram (25
ms Hanning window, 10 ms step) context windows represented as $F\!=\!64$ mel
bins by $T\!=\!96$ STFT frames (for a duration of 0.96 s).  Due to its proven
success in past audio modeling
research~\cite{hershey2016,jansen2018unsupervised}, we use the ResNet-50
architecture for the audio embedding networks ($f$), with the final average pool
followed by a $d\!=\!128$-dimensional embedding layer.  For simplicity we use
the identical architecture for the image network ($g$), changing only the input
size to match the image dimensions.

Both coincidence prediction networks, $p_{\mathrm{AA}}$ and $p_{\mathrm{AV}}$,
are defined by a single 512-unit fully connected ReLU layer followed by a binary
classification output layer (i.e., logistic regression).  Each cluster network
($p_{\mathrm{clust}}$) is a single fully connected layer with $K$ outputs,
followed by a softmax nonlinearity to produce probabilities.  To improve
compatibility with our learned embeddings, which are amenable to cosine
distance, we follow~\cite{arora2019deep} by length normalizing both input
embeddings and layer weights for each output logit and introduce a fixed logit
scaling factor of 60 before applying the softmax.  We use diversity
hyperparameter $\gamma=1.1$ for all clustering models and learning curriculum
hyperparameters $\alpha=\beta=0.1$ for all experiments so that stronger
self-supervised learning criteria dominate. (Note that downstream evaluation
performance is not highly sensitive to these settings within reasonable ranges).
We evaluate our proposed learning framework on both unsupervised audio
representation learning and lightly-supervised classification tasks, described
in turn below.

\subsection{Unsupervised Audio Representation Learning}

To measure the utility of our unsupervised representations, we reuse the
query-by-example (QbE) and shallow classifier evaluation methodology
of~\cite{jansen2018unsupervised}.  This involves training all unsupervised
models on the entirety of the AudioSet training set, while ignoring the
labels. Performance is characterized relative to the raw spectrogram features
(baseline) and the fully-supervised triplet embedding (topline)
from~\cite{jansen2018unsupervised}, reporting each unsupervised method's
recovery of that range.

The first evaluation is a QbE semantic retrieval task, which directly measures
the utility of the distance implied by the learned embedding space.  For each
class, we compute pairwise cosine distances between a set of positive and
negative clips for each class and sort them by ascending distance.  We then
characterize retrieval performance with the average precision for ranking
within-class pairs higher.  The second evaluation is classification with a
simple architecture, where the entire labeled AudioSet is used to train a
527-way classifier with one 512-unit hidden layer taking the fixed embedding as
input.  While this simple classifier is fully supervised, it benchmarks the
utility of each embedding model as a fixed representation for downstream
supervised semantic classification tasks.

Table~\ref{tab:comb1} shows the performance for the (a-b) baseline and topline;
(c) the temporal proximity-based unsupervised triplet embedding
from~\cite{jansen2018unsupervised}, which is a state-of-the-art audio-only
unsupervised embedding technique; (d) our implementation of the AV
correspondence audio embedding~\cite{arandjelovic2017look}, where we follow the
original recipe that uses VGG architecture, random negative sampling, and
$\Delta T = 1$ s; (e-f) the AA and AV coincidence embeddings that use objectives
$\mathcal{L}_{\mathrm{AA}}$ and $\mathcal{L}_{\mathrm{AV}}$ separately, along
with ResNet-50 and $\Delta T = 10 s$; (g-i) AV coincidence ablation experiments
to characterize our changes to the original AV correspondence recipe; and (j-k)
the joint AV+AA coincidence loss, both with ($\mathcal{L}_{\mathrm{joint}}$) and
without ($\mathcal{L}_{\mathrm{coin}}$) the cluster loss (in this case using
$K$=1 million output clusters).  

AA coincidence matches the earlier temporal proximity triplet approach for QbE,
though there is a moderate loss for shallow model training. The AV
correspondence recipe from~\cite{arandjelovic2017look} gives large improvements
over all audio-only models (absolute 20\% and 9\% recovery for QbE and
classification, respectively), confirming the power of cross-modal
self-supervision demonstrated in that previous work. However, our
generalization to AV coincidence provides substantial improvements over the AV
correspondence recipe (12\% and 9\% absolute recovery gain), with both the
ResNet-50 upgrade and all-pairs batch construction providing lift in one or both
tasks.  The increase of coincidence time scale to $\Delta T=10$ s performs
equivalently to using overlapping AV frames.  This indicates the constraint of
direct correspondence proposed in~\cite{arandjelovic2017look} is unnecessary for
semantic cross-modal AV learning, allowing us to unify the time scale with AA
coincidence, which requires longer time scales for success.  We find that joint
training provides additional gains: the coincidence and clustering objective
combination more than doubles the audio-only model recovery for QbE while nearly
matching the fully supervised triplet model as a representation for downstream
supervised classification tasks.

\subsection{Sound Classification with Active Learning}

Next we evaluate the cluster-based active learning procedure introduced in
Section~\ref{sec:consolidation}.  We simulate both random labeling baselines and
active learning procedures using the AudioSet labels.  To adequately simulate
the proposed cluster labeling procedure, we must reduce to a 115-class, mutually
exclusive subset of AudioSet ontology for the remaining experiments, which
guarantees all examples are fully annotated. However, since the labels apply at
the clip-level, restricting to this class subset will still bring along a
substantial amount of out-of-set audio, making our simulation a worst-case
approximation to the real problem. 

\begin{table}[t]
\vspace{-0.2cm}
  \centering
  \caption{Clustering performance and corresponding cluster-based label quality.}
  \label{tab:comb2}
  \vspace{0.2cm}
  \begin{tabular}{c|c|c|c|c|c}
    \hline
    {\bf $K$} & {\bf \# Active} & {\bf VM} & {\bf \# Labeled} & {\bf Recall} & {\bf Precision} \\
    \hline
    1K   & 968    & 0.553 & 370    & 0.269 & 0.097 \\ 
    10K  & 7,575  & 0.639 & 3,700  & 0.417 & 0.118 \\        
    100K & 35,830 & 0.668 & 35,830 & 0.560 & 0.109 \\
    1M   & 77,614 & 0.674 & 37,000 & 0.549 & 0.117 \\
    \hline
  \end{tabular}
\end{table}

We first measure intrinsic performance of the clustering method presented in
Section~\ref{sec:categorization}. Table~\ref{tab:comb2} shows the context-window
clustering performance in terms of V-Measure (VM)~\cite{rosenberg2007v}, a
standard clustering metric, along with the corresponding label precision and
recall resulting from the cluster-based labeling procedure.  Since the
clustering model is trained on the entirety of AudioSet but evaluated on the
mutually exclusive 115-class subset that excludes the high-prior speech and
music classes, growth in active clusters is sublinear in $K$.  We also see that
VM plateaus, indicating the additional finer grained partitioning is not focused
on the 115 target classes but instead the background sounds in the evaluation
segments.  As we increase $K$ and correspondingly increase the number of
clusters labeled, we observe marked improvements to the label recall for a
roughly fixed precision (which is limited by the clip-level nature of AudioSet
labels).  By labeling 37K examples, each representing a distinct cluster, we are
able to recover on average half of the positive example labels for the 115-class
set with an amount of label noise that is roughly in line with the weakness of
the clip-level labels we are using (i.e., for many classes, the sound event only
occupies a portion of the AudioSet clip, which means the example-level ground
truth labels we score against are inherently noisy).

\begin{table}[t]
\vspace{-0.2cm}
  \centering
  \caption{Classifier performance for random and cluster labeling.}
  \label{tab:finalclass}
  \vspace{0.2cm}
  \begin{tabular}{c|c|c|c|c}
    \hline
                         & {\bf Label}  & {\bf Examples w/}  &           &          \\
    {\bf Label Strategy} & {\bf Budget} & {\bf Labels} & {\bf mAP} & {\bf d\textprime} \\
    \hline
    Complete (topline)      & 3.7M  & 3.7M & 0.566 & 2.58 \\
    \hline
    \multirow{5}{*}{Random} & 370M   & 370K & 0.421 & 2.28 \\
                            & 185M   & 185K & 0.350 & 1.96 \\
                            & 74K    & 74K  & 0.246 & 1.71 \\
                            & 37K    & 37K  & 0.211 & 1.56 \\
                            & 3.7K   & 3.7K & 0.083 & 1.00 \\
                            & 370    & 370  & 0.028 & 0.44 \\
    \hline
    Cluster:$K=1$M          & 37K    & 3.3M & 0.335 & 2.15 \\
    Cluster:$K=10$K         & 3.7K   & 3.0M & 0.267 & 1.80 \\
    Cluster:$K=1$K          & 370    & 2.7M & 0.150 & 1.10 \\
    \hline
  \end{tabular}
\end{table}

Next, we use these cluster-based labels to train classifiers.
Table~\ref{tab:finalclass} shows the 115-way classifier performance for random
example sampling baselines (trained from scratch) and the proposed cluster-based
labeling procedure (trained using the curriculum defined in
Section~\ref{sec:curriculum}).  Here, we vary annotation budget and measure the
resulting clip-level classifier performance (we average 0.96 second frame-level
scores for each class across the clip to arrive at the clip-level score, as done
in~\cite{audioset}) in terms of mean average precision (mAP) and mean
{d\textprime} (see~\cite{audioset} for details on this useful, but nonstandard,
evaluation metric).  With a small budget and skewed class prior distributions,
random sampling tends to overspend on a common classes, yielding little
supervision for rare classes.\footnote{Note that while we train with very few
  labeled examples for some random sampling baselines, the ResNet-50
  architecture still outperforms smaller architectures, a finding consistent
  with~\cite{jansen2018unsupervised}.}  Guiding the annotator effort via the
discovered categorical structure, we can turn as little as 370 annotations into
cluster-based labels for well over half the examples in the dataset with
improved class coverage.  Despite the label noise from cluster label
propagation, the improvements over random sampling are dramatic; for fixed
annotation budget, mean average precision increases by up to a factor of 5.3x,
while {d\textprime} increases by up to 0.8 absolute.  If targeting a specific
classifier performance, our cluster-based labeling procedure reduces the
required label budget by approximately 2-20x over the range we evaluated.

\section{Conclusions}
\label{sec:conclusion}

We have presented and evaluated a novel learning framework for building sound
representation and recognition models with only a small number of labeled
examples.  We demonstrated that, by unifying past self-supervised learning
successes into a generalized notion of coincidence prediction, we can learn
powerful semantic audio representations that enable a highly efficient use of a
given annotation budget.  By introducing a neural clustering objective, we can
simultaneously partition the space for cluster-based active learning while also
improving the semantic structure of the embedding space itself, leading to a new
high water mark for unsupervised audio representation learning.  Moreover, the
cluster-based annotations amplify the impact of each solicited label, producing
dramatic classification performance gains over random sampling.

\section{Acknowledgments}

The authors thank Elad Eban, Malcolm Slaney, and Matt Harvey of Google Research
for helpful discussions and feedback.

\bibliographystyle{IEEEbib}
\bibliography{arxiv_version.bib}

\end{document}